\documentclass[aps,superscriptaddress,eqsecnum,tightenlines,preprint,nofootinbib]{revtex4}
\usepackage{graphicx,amssymb,amsbsy,bm,amsmath}

\usepackage{hyperref}

\newcommand{\vx}{\ensuremath{\vec{x}}}

\newcommand{\vy}{\ensuremath{\vec{y}}}
\newcommand{\vk}{\ensuremath{\vec{k}}}
\newcommand{\vq}{\ensuremath{\vec{q}}}
\newcommand{\vp}{\ensuremath{\vec{p}}}

\newcommand{\be}{\begin{equation}}
\newcommand{\ee}{\end{equation}}
\newcommand{\bea}{\begin{eqnarray}}
\newcommand{\eea}{\end{eqnarray}}

\begin{document}
\title{An effective field theory during inflation: \\ reduced density matrix and its  quantum master equation.}
\author{D. Boyanovsky}
\email{boyan@pitt.edu} \affiliation{Department
of Physics and Astronomy,\\ University of Pittsburgh\\Pittsburgh,
Pennsylvania 15260, USA}
\date{\today}
\begin{abstract}
We study the power spectrum of super-Hubble fluctuations of an inflaton-like scalar field, the ``system'', coupled to another scalar field, the ``environment'' during de Sitter inflation. We obtain the reduced density matrix for the inflaton fluctuations by integrating out the environmental degrees of freedom. These are  considered to be massless and conformally coupled to gravity as a \emph{proxy} to describe degrees of freedom that remain sub-Hubble all throughout inflation. The time evolution of the density matrix is described by a quantum master equation, which describes the decay of the vacuum state, the production of particles and correlated pairs and quantum entanglement between super and sub-Hubble degrees of freedom. The quantum master equation  provides a non-perturbative resummation of secular terms from self-energy (loop) corrections to the inflaton fluctuations. In the case studied here these are Sudakov-type double logarithms which result in the \emph{decay} of the power spectrum of inflaton fluctuations upon horizon crossing with a concomitant violation of scale invariance. The reduced density matrix and its quantum master equation furnish a powerful non-perturbative framework to study the effective field theory of long wavelength fluctuations by tracing short wavelength degrees of freedom.
\end{abstract}


\maketitle

\section{Introduction}\label{sec:intro}
Cosmology is entering the precision era, measurements of the cosmic microwave background anisotropies by the WMAP\cite{wmap} and PLANCK\cite{planck} missions have provided confirmation of a nearly scale invariant power spectrum of adiabatic perturbations, one of the main predictions of inflation. These measurements support the main paradigm of inflation based on a   scalar field slowly rolling down a potential landscape leading to a nearly de Sitter inflationary stage   and the generation of perturbations from the quantum fluctuations that are amplified when their wavelengths cross the Hubble radius. This simple and compelling   picture  invites deeper scrutiny at a fundamental level. Most models of inflation rely on either a single scalar field or several scalar fields (which typically yield isocurvature  or entropy  perturbations) but no interactions between the inflaton field and the large number of degrees of freedom of the standard model of particle physics (and beyond). Excitation of the degrees of freedom that would populate a radiation dominated era post inflation is \emph{assumed} to happen at the end of inflation during a period of ``reheating''\cite{reheat1,drewes,reheatrev}, however, this necessarily implies a \emph{coupling} between the inflaton   and the degrees of freedom that describe the physics of the radiation dominated era. Such   interaction between the inflaton and other (fermionic,   scalar) degrees of freedom cannot just switch-on at the end of inflation, and on physical grounds   should be expected to be present even during the inflationary stage.
Interactions of quantum fields in de Sitter (or nearly de Sitter) space-time have been the focus of several studies\cite{woodardcosmo,proko1,decayds,akhmedov,woodard1,prokowood,onemli,sloth1,sloth2,riotto,fermionswoodpro,picon,lello,rich,raja,
richboy,boyan,serreau,parentani,smit} which show strong infrared and secular effects and the possibility that the vacuum state in de Sitter space time may be unstable towards decay\cite{polyakov}. Non-Gaussianity is a manifestation of self-interactions of curvature perturbations and could leave an observable imprint on the cosmic microwave background, although it is  argued to be small in    single field slow roll inflationary models\cite{maldacena,komatsu,weinberg}.

Interactions with heavy fields with masses $M\gg H$ with $H$ the Hubble parameter during inflation have been treated in terms of effective field theory descriptions\cite{heavyachu,heavycespe,heavy1,heavy2} mainly by neglecting kinetic terms and correlations,   effectively treating the heavy degrees of freedom as auxiliary fields that can be ``integrated out'' at tree level, or including correlations of the heavy fields in powers of $H/M \ll 1$\cite{heavyjack}.

In a non-equilibrium situation as is the case with cosmological expansion,   integrating out   short wavelength degrees of freedom leads to the effective field theory description in terms of a \emph{reduced density matrix} for long wavelength fluctuations. Such a description is, fundamentally, akin to a \emph{Wilsonian} approach to an effective field theory\cite{bala} by coarse graining short distance degrees of freedom.  At the level of a non-equilibrium effective action, the study of the effects of tracing out degrees of freedom  was pioneered with the study of quantum Brownian motion\cite{feyn,schwinger,calzetta,paz}, the degrees of freedom of interest are considered to be the ``system'' whereas those that are integrated out (traced over) are the ``bath'' or ``environment''. The effects of the bath or environment are manifest in the non-equilibrium effective action via an \emph{influence action} which is in general non-local and describes dissipative processes. This influence action is   determined by the \emph{correlation functions} of the environmental degrees of freedom,  and determines the time evolution of the reduced density matrix. An alternative but equivalent description of the time evolution of the reduced density matrix is the \emph{quantum master equation}\cite{breuer,zoeller} which includes the effects of coupling to the environmental degrees of freedom via their quantum mechanical correlations.

 A generic quantum master equation approach   for a reduced density matrix describing cosmological perturbations has been advocated in ref.\cite{burhol} in terms of local correlations of environmental degrees of freedom.

In ref. \cite{boyeff} the equivalence between the influence functional  and the  quantum master equation in Minkowski space-time  was established,  and shown that they provide a non-perturbative resummation of self-energy diagrams directly in real time providing an effective  field theory description of non-equilibrium phenomena.

\vspace{2mm}

\textbf{Motivations and goals:} In this article we consider an inflaton-like scalar field as ``the system'' in interaction with other fields considered as the ``environment''  with the goal of studying the influence of sub-Hubble degrees of freedom  of the ``environment''  upon the power spectrum of super-Hubble fluctuations of  the ``system''  during de Sitter inflation. The quantum fluctuations of the inflaton-like scalar (the system)  are amplified and become classical when their wavelengths become larger than the Hubble radius during inflation  and in the non-interacting theory their power spectrum becomes nearly scale invariant.     We study the influence of the environmental fields  by obtaining the reduced density matrix for  the inflaton-like scalar field,   and the quantum master equation that describes its time evolution by consistently tracing over the environmental degrees of freedom.  Our goal is to obtain the corrections to the power spectrum of super-Hubble fluctuations of the inflaton-like scalar from the interaction with degrees of freedom whose quantum fluctuations remain sub-Hubble all throughout the inflationary stage. For this purpose we consider the system to be a minimally coupled scalar (inflaton-like)    coupled to a massless \emph{conformally} coupled scalar field--the ``environment''-- that serves as a \emph{proxy} for fields (including fermionic fields) whose quantum correlations   do not become amplified for super-Hubble wavelengths. We also comment on the case of minimally coupled environmental fields.  The mode functions of the quantum environmental fields effectively describe quantum fluctuations that remain  sub-Hubble all throughout inflation.

\vspace{2mm}

\textbf{Brief summary of results:} We obtain the reduced density matrix   of the inflaton-like scalar field (the system)  and its quantum master equation, by tracing out sub-Hubble degrees of freedom (the ``environment'') up to second order in the coupling.  We consider the case in which  both the ``system'' and the ``environment'' are in their Bunch-Davies vacuum states at the beginning of the inflationary stage.      A perturbative analysis of the quantum master equation explicitly shows the decay of the vacuum state and the  production of single particles as well as correlated pairs which lead to \emph{quantum entanglement} between the inflaton fluctuations and those of the environmental fields. The full solution of the quantum master equation provides a non-perturbative resummation of self-energy diagrams   determined by the correlation functions (loops) of the environmental degrees of freedom.  From the quantum master equation we obtain the equations of motion for super-Hubble correlations of the inflaton field from which we extract the power spectrum. Its solution provides a non-perturbative resummation of secular Sudakov-type double logarithms from the inflaton self energy     and yields the corrections to the power spectrum of super-Hubble fluctuations. These indicate the \emph{decay} of the power spectrum after ``horizon crossing'' and violation of scale invariance even when the power spectrum in absence of interactions is scale invariant.

\section{The model:}\label{model}
We consider a spatially flat Friedmann-Robertson-Walker (FRW)
cosmological space-time and two interacting scalar fields $\phi,\varphi$ although the methods and broad conclusions will be more general.  The field $\phi$ is an inflaton-like scalar field minimally coupled to gravity, this is the ``system'', and the field $\varphi$  is the ``environment'' as discussed below it will be chosen to be a massless, conformally coupled scalar field and it will be traced out of the total density matrix to yield the reduced density matrix for $\phi$.

 In comoving
coordinates, the action is given by
\bea
S & = &\int d^3x \; dt \;  a^3(t) \Bigg\{
\frac{1}{2}{\dot{\phi}^2}-\frac{(\nabla
\phi)^2}{2a^2}-\frac{1}{2}\Big(M^2_\phi+\xi_\phi \; \mathcal{R}\Big)\phi^2  \nonumber \\ & + &
\frac{1}{2}{\dot{\varphi}^2}-\frac{(\nabla
\varphi)^2}{2a^2}-\frac{1}{2}\Big(M^2_\varphi+\xi_\varphi \; \mathcal{R}\Big)\varphi^2   -  \lambda \,  \phi\,:\varphi^2:  \Bigg\}\,, \label{lagrads}
\eea
with \be \mathcal{R} = 6 \left(
\frac{\ddot{a}}{a}+\frac{\dot{a}^2}{a^2}\right) \ee being the Ricci
scalar,   $\xi=
0,1/6$ correspond  to minimal and
conformal coupling respectively.  The interaction has been normal-ordered
\be :\varphi^2: = \varphi^2- \langle \varphi^2 \rangle \label{no} \ee where the brackets $\langle (\cdots)\rangle$ refer to the expectation value in the initial density matrix (see below).

In the case of  de Sitter space time with $a(t) = e^{H t}$, it is convenient to pass to conformal time $\eta = -e^{-Ht}/H$ with
\be  a(t(\eta)) = -\frac{1}{H\eta} \,,\label{aofeta}\ee
  and introduce a conformal rescaling of the fields
\begin{equation}
 \phi(\vx,t) = \frac{\chi(\vx,\eta)}{a(t(\eta))}~~;~~ \varphi(\vx,t) = \frac{\psi(\vx,\eta)}{a(t(\eta))}.\label{rescale}
\end{equation}
After discarding surface terms the action becomes   \be S  =
  \int d^3x \; d\eta  \; \Bigg\{\frac12\left[
{\chi'}^2-(\nabla \chi)^2-\mathcal{M}^2_{\chi}(\eta) \; \chi^2 +
{\psi'}^2-(\nabla \psi)^2-\mathcal{M}^2_{\psi}(\eta) \; \psi^2  \right] +\frac{\lambda}{H\eta} \;  \chi\,:\psi^2:   \Bigg\} \; , \label{rescalagds}\ee
with primes denoting derivatives with respect to
conformal time $\eta$ and
\be
\mathcal{M}^2_{\chi,\psi}(\eta)  = \Big[\frac{M^2_{\chi,\psi}}{H^2}+12\Big(\xi_{\chi,\psi} -
\frac{1}{6}\Big)\Big]\frac{1}{\eta^2}   \; , \label{massds2}
\ee where for consistency of notation we have called
\be M_{\phi,\varphi} \rightarrow M_{\chi,\psi}~~;~~ \xi_{\phi,\varphi}\rightarrow \xi_{\chi,\psi}\,  \label{redefs}\ee respectively.

Since $\lambda$ has dimensions of mass we will consider the weak coupling case with $\lambda/H\ll 1$. In the non-interacting case $\lambda =0$ the Heisenberg equations of motion for the spatial Fourier modes of wavevector $\vec{k}$ for the conformally rescaled fields are
\bea
&& \chi''_{\vk}(\eta)+
\Big[k^2-\frac{1}{\eta^2}\Big(\nu^2_\chi -\frac{1}{4} \Big)
\Big]\chi_{\vk}(\eta)  =   0    \\
&& \psi''_{\vk}(\eta)+
\Big[k^2-\frac{1}{\eta^2}\Big(\nu^2_\psi -\frac{1}{4} \Big)
\Big]\psi_{\vk}(\eta)  =   0  \label{modes}\eea
where
\be \nu^2_{a} = \frac{9}{4}- \Big(\frac{M^2_{a}}{H^2}+12\, \xi_a \Big) ~~;~~ a= \chi,\psi \,.
\label{nusa} \ee

The Heisenberg fields are expanded in a comoving volume $V$ as
\bea
\chi(\vx,\eta) & = & \frac{1}{\sqrt{V}}\,\sum_{\vq} \Big[b_{\vq}\,g(q,\eta)+ b^\dagger_{-\vq}\,g^*(q,\eta) \Big]\,e^{i\vq\cdot\vx} \label{chiex} \\
\psi(\vx,\eta) & = & \frac{1}{\sqrt{V}}\,\sum_{\vk} \Big[a_{\vq}\,u(q,\eta)+ a^\dagger_{-\vq}\,u^*(q,\eta) \Big]\,e^{i\vk\cdot\vx}\,. \label{psiex} \eea
We choose Bunch-Davies conditions in both fields, namely
\be b_{\vq} |0\rangle_{\chi} =0 ~~;~~ a_{\vk} |0\rangle_{\psi} =0 \label{bdvac}\ee
and
\bea
 && g(q,\eta)= \frac{1}{2}\,e^{i\frac{\pi}{2}(\nu_\chi+\frac{1}{2})}\,\sqrt{-\pi\,\eta}\,H^{(1)}_{\nu_\chi}(-q\eta)\label{gqeta}\\&&
 u(k,\eta)= \frac{1}{2}\,e^{i\frac{\pi}{2}(\nu_\psi+\frac{1}{2})}\,\sqrt{-\pi\,\eta}\,H^{(1)}_{\nu_\psi}(-k\eta)\,.\label{uketa}\eea These conditions may be generalized to non-Bunch-Davies, but here we consider this simpler case to highlight the main physical consequences.

 The $\chi$ field is considered to be minimally coupled, $\xi_\chi =0$ and nearly massless with $M_\chi/H \ll1$, from which it follows that as $-q\eta \rightarrow 0$
 \be g(q,\eta) \propto 1/\eta \,.\label{smaleta}\ee This behavior in the super-Hubble limit will lead to strong secular contributions in the long time limit.

 The time evolution of a   density matrix initially prepared at time $\eta_0$ is given by
 \be \rho(\eta)= U(\eta,\eta_0)\,\rho(\eta_0)\,U^{-1}(\eta,\eta_0) \,,\label{rhoeta}\ee where $\mathrm{Tr}[\rho(\eta_0)]=1$ and  $U(\eta,\eta_0)$ is the unitary time evolution of the full theory, it obeys
\be i\frac{d}{d\eta} U(\eta,\eta_0) = H(\eta) \,U(\eta,\eta_0)~~;~~ U(\eta_0,\eta_0) =1 \label{U} \ee where $H(\eta)$ is the total Hamiltonian. Writing the total Hamiltonian in terms of the free and interaction Hamiltonians as $H(\eta)= H_0(\eta)+H_i(\eta)$ it is convenient to pass to the interaction picture introducing the unitary time evolution operator of the free theory $U_0(\eta,\eta_0)$ obeying
 \be i\frac{d}{d\eta} U_0(\eta,\eta_0) = H_0(\eta) \,U_0(\eta,\eta_0)~~;~~ U_0(\eta_0,\eta_0) =1 \label{Uzero} \ee and define the density matrix in the interaction picture
 \be \rho_I(\eta) = U^{-1}_0(\eta,\eta_0) \rho(\eta) U_0(\eta,\eta_0) \equiv U_I(\eta,\eta_0) \rho(\eta_0) U^{-1}_I(\eta,\eta_0) \label{rhoI}\ee where $U_I(\eta,\eta_0)$ is the unitary time evolution operator in the interaction picture obeying
 \be  i\frac{d}{d\eta} U_I(\eta,\eta_0) = H_I(\eta) \,U_I(\eta,\eta_0)~~;~~ U_I(\eta_0,\eta_0) =1 \label{UI}\ee where
 \be H_I(\eta) = U^{-1}_0(\eta,\eta_0) H_i(\eta) U_0(\eta,\eta_0)\,. \label{Hintpic}\ee  From (\ref{lagrads}) we find
 \be H_I(\eta) = -\frac{\lambda}{H\eta}\int d^3 x \,\chi(\vx,\eta)\,:\psi^2(\vx,\eta): \label{HIeta}\ee and the $\chi,\psi$ fields are in the free field Heisenberg representation (\ref{chiex},\ref{psiex}). Normal ordering is defined as
 \be :\psi^2(\vx,\eta): = \psi^2(\vx,\eta) - \mathrm{Tr}[\psi^2(\vx,\eta)\rho_I(\eta_0)] \label{norord}\ee

 \section{Quantum Master Equation}\label{sec:qme}
 The steps leading to the quantum master equation up to second order in the coupling are given in detail in Minkowski space time in ref.\cite{boyeff}. In this reference the equivalence between integrating out the heavy (or short wavelength) degrees of freedom in the path integral representation and the quantum master equation is established up to second order in the coupling. This equivalence and the relation to a stochastic description  translate directly to the case of an FRW cosmology, these more formal aspects are relegated to a companion article\cite{boyeffnext}. In this article we obtain the quantum master equation directly and apply it to understand several physical consequences.

 The time evolution of the density matrix in the interaction picture is given by
 \be  \rho'_I(\eta)= -i [H_I(\eta),\rho_I(\eta)]\, \label{detarho}\ee whose formal solution is \be \rho_I(\eta)= \rho_I(\eta_0) -i \int^{\eta}_{\eta_0} \,[H_I(\eta'),\rho_I(\eta')]\,d\eta' \,, \label{solurho}\ee this solution is inserted back into (\ref{detarho}) leading to the iterative equation
 \be \rho'_I(\eta) = -i[H_I(\eta),\rho_I(\eta_0)] - \int^{\eta}_{\eta_0} \,[H_I(\eta),[H_I(\eta'),\rho_I(\eta')]]\,d\eta' \,.\label{itersolu}\ee
 The next steps leading to the quantum master equation rely on various approximations, discussed in detail in \cite{boyeff}. The first is \textbf{factorization}, namely
  \be \rho_I(\eta) = \rho_{I\chi}(\eta)\otimes \rho_{I\psi}(\eta_0) \label{factorization} \ee
this is an ubiquitous approximation\cite{breuer,zoeller}. In ref.\cite{boyeff} it is shown that this approximation results from obtaining the non-equilibrium effective action in the path integral  Schwinger-Keldysh formulation   in a consistent cumulant expansion of the trace over the  environmental  degrees of freedom when the initial density matrix is factorized. These aspects  translate directly to an FRW cosmology in conformal time as discussed in a companion article\cite{boyeffnext}. In this article we \emph{assume} (as is common in the literature) that the initial density matrix at the beginning of inflation is of the factorized form, it is clearly interesting to consider the case of initial correlations, which is postponed to further study.

Taking the trace of $\rho_I(\eta)$ over the $\psi$ degrees of freedom yields the \emph{reduced density matrix}
\be \rho_r(\eta) = {\mathrm{Tr}}_{\{\psi\}}\, \rho_I(\eta)\,. \label{rhored}\ee   The normal ordering (\ref{norord}) entails that the first term in (\ref{itersolu}) vanishes upon taking the trace over $\psi$ leading to
\bea
\rho'_r(\eta) & = & -\frac{\lambda^2}{H^2\,\eta} \int^\eta_{\eta_0}\frac{d\eta'}{\eta'}\,\int d^3x \,\int d^3 y \Bigg\{ \chi(\vx,\eta)\chi(\vy,\eta') \rho_r(\eta')\,G^>(\vx-\vy,\eta,\eta')  \nonumber \\
&+& \rho_r(\eta') \chi(\vy,\eta')  \chi(\vx,\eta) \,G^<(\vx-\vy,\eta,\eta')- \chi(\vx,\eta)\rho_r(\eta')\chi(\vy,\eta')\,G^<(\vx-\vy,\eta,\eta') \nonumber \\ & - &
\chi(\vy,\eta') \rho_r(\eta') \chi(\vx,\eta)\,G^>(\vx-\vy,\eta,\eta')\Bigg\} \,,\label{rhoreddot}\eea
where
\bea && G^>(\vx-\vy,\eta,\eta') = \mathrm{Tr}[:\psi^2(\vx,\eta):\,:\psi^2({\vy},\eta'):\,\rho_{I\psi}(\eta_0)] \,,\label{ggreat} \\
&& G^<(\vx-\vy,\eta,\eta') = \mathrm{Tr}[:\psi^2(\vy,\eta'):\,:\psi^2(\vx,\eta):\,\rho_{I\psi}(\eta_0)]\,. \label{gless} \eea In writing the correlation functions $G^{>,<}$ as functions of $\vx-\vy$ we used spatial translational invariance in a spatially flat FRW space-time which allows us to write,
\be G^>(\vx-\vy,\eta,\eta') = \frac{1}{V}\sum_{\vp} \mathcal{K}^>[p,\eta,\eta']\, e^{-i\vp\cdot(\vx-\vy)}~~;~~ G^<(\vx-\vy,\eta,\eta') = \frac{1}{V}\sum_{\vp} \mathcal{K}^<[p,\eta,\eta']\, e^{-i\vp\cdot(\vx-\vy)}\,,  \label{ks}\ee where $V$ is the quantization volume.

A this stage, the  second   \textbf{Markov} approximation is invoked:  taking $\rho_r(\eta') \rightarrow \rho_r(\eta)$. This  is justified in weak coupling:  since $d \rho_r(\eta)/d \eta \propto  \lambda^2/H^2 \ll 1$, an integration by parts\cite{boyeff} and neglecting contributions of $\mathcal{O}(\lambda^4)$ on the right hand side of (\ref{rhoreddot}) leads to the Markov approximation. The main arguments showing the validity of this approximation for weak coupling are available in ref.\cite{boyeff}.

\begin{figure}[ht!]
\begin{center}
\includegraphics[height=3.5in,width=3.5in,keepaspectratio=true]{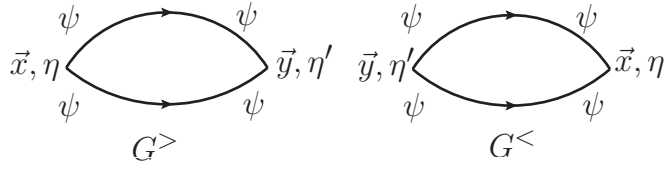}
\caption{The correlation functions $G^>(\vx-\vy,\eta,\eta'),G^<(\vx-\vy,\eta,\eta') $. }
\label{fig:correlators}
\end{center}
\end{figure}

In order to obtain the correlation functions $G^{<,>}$  we need to specify $\rho_{I\psi}(\eta_0)$, we consider the simple case
\be \rho_{I\psi}(\eta_0) = |0\rangle_{\psi} {}_{\psi}\langle 0| \label{rhopsi0}\ee where $|0\rangle_\psi$ is the Bunch-Davies vacuum for the $\psi$ fields and $-\eta_0$ is taken to be the beginning of the (nearly) de Sitter inflationary stage, although a generalization is   straightforward. In this case we find
\bea && \mathcal{K}^>[q;\eta,\eta'] \equiv K[q;\eta,\eta'] = 2 \int \frac{d^3k}{(2\pi)^3}\,u(k,\eta)u^*(k,\eta')u(p,\eta)u^*(p,\eta')~~;~~ p=|\vk+\vq|\nonumber \\
&& \mathcal{K}^<(q;\eta,\eta')= \mathcal{K}^>[q;\eta',\eta]= K^*[q;\eta,\eta'] \,,\label{kernels} \eea where $u(k,\eta)$ etc, are the mode functions (\ref{uketa}). The Feynman diagrams for the correlation functions $G^>,G^<$ are shown in fig.(\ref{fig:correlators}).

Using the expansion (\ref{chiex}) we find the general form of the quantum master equation up to second order in the interaction (\ref{HIeta})
\bea
\rho'_r(\eta) & = & -\frac{\lambda^2}{H^2\,\eta} \int^\eta_{\eta_0}\frac{d\eta'}{\eta'}\,\sum_{\vq} \Bigg\{ \chi_{\vq}(\eta)\chi_{-\vq}(\eta') \rho_r(\eta)\, {K}(q;\eta,\eta')   + \rho_r(\eta) \chi_{-\vq}(\eta')  \chi_{\vq}(\eta) \,{K}^*(q;\eta,\eta') \nonumber \\ &-&  \chi_{\vq}(\eta)\rho_r(\eta)\chi_{-\vq}(\eta')\, {K}^*(q;\eta,\eta')   -
\chi_{-\vq}(\eta') \rho_r(\eta) \chi_{\vq}(\eta)\, {K}(q;\eta,\eta')\Bigg\} \label{qme1}\eea where
\be \chi_{\vq}(\eta)= b_{\vq}\,g(q,\eta)+ b^\dagger_{-\vq}\,g^*(q,\eta) ~~;~~ \chi_{-\vq}(\eta) = \chi^\dagger_{\vq}(\eta) \,.\label{chiofq} \ee

To obtain expectation values of operators in general one has to solve the quantum master equation and find $\rho_r(\eta)$. However, it is most useful to obtain the evolution equations for operators that do not evolve in time in the interaction picture. Consider one such operator $\mathcal{O}$ which is constant in the interaction picture, then with $\langle \mathcal{O} \rangle(\eta) = \mathrm{Tr}\mathcal{O}\rho_r(\eta)$ it follows that
\be \frac{d}{d\eta} \langle \mathcal{O} \rangle(\eta) = \mathrm{Tr}\Big( \mathcal{O} \,\rho'_r(\eta)\Big)\,. \label{evolOp}\ee

It proves illuminating to write the quantum master equation in terms of the operators $b^\dagger_q, b_q$, this yields simpler expressions for the matrix elements of the reduced density matrix in the Fock basis of quanta associated with these operators, in the case under consideration these are Bunch-Davies Fock states. We find
\bea \rho'_r(\eta)   = &&  -\frac{\lambda^2}{H^2\,\eta} \int^\eta_{\eta_0}\frac{d\eta'}{\eta'}  \sum_{\vq}   \Bigg\{ g(q,\eta')g^*(q,\eta) K[q,\eta,\eta']\Big[b^\dagger_{\vq} b_{\vq} \,\rho_r(\eta) - b_{\vq}\, \rho_r(\eta)\, b^\dagger_{\vq} \Big]\nonumber \\ & + & g(q,\eta)g^*(q,\eta') K^*[q;\eta,\eta']\Big[\rho_r(\eta\,)b^\dagger_{\vq} b_{\vq}   - b_{\vq}\, \rho_r(\eta) \, b^\dagger_{\vq} \Big]   \nonumber \\ & + & g(q,\eta)g^*(q,\eta')K[q;\eta,\eta']\Big[b_{\vq} b^\dagger_{\vq}\,\rho_r(\eta)-b^\dagger_{\vq}\,\rho_r(\eta)\,b_{\vq} \Big] \nonumber \\ & + &  g(q,\eta')g^*(q,\eta)K^*[q;\eta,\eta']\Big[\rho_r(\eta)\,b_{\vq} b^\dagger_{\vq} -b^\dagger_{\vq}\,\rho_r(\eta)\,b_{\vq} \Big] \nonumber \\ & + & g(q,\eta)g(q,\eta')K[q;\eta,\eta']\Big[b_{\vq}b_{-\vq} \,\rho_r(\eta)- b_{-\vq}\,\rho_r(\eta)\,b_{\vq} \Big] \nonumber \\ & + &
g(q,\eta)g(q,\eta')K^*[q;\eta,\eta']\Big[\rho_r(\eta)\,b_{\vq}b_{-\vq} - b_{-\vq}\,\rho_r(\eta)\,b_{\vq} \Big] \nonumber \\ & + &
g^*(q,\eta)g^*(q,\eta')K[q;\eta,\eta']\Big[b^\dagger_{\vq}b^\dagger_{-\vq} \,\rho_r(\eta)- b^\dagger_{-\vq}\,\rho_r(\eta)\,b^\dagger_{\vq} \Big] \nonumber \\ & + &
g^*(q,\eta)g^*(q,\eta')K^*[q;\eta,\eta']\Big[\rho_r(\eta)\,b^\dagger_{\vq}b^\dagger_{-\vq} - b^\dagger_{-\vq}\,\rho_r(\eta)\,b^\dagger_{\vq} \Big]\Bigg\}\,. \label{bbdagrho}
\eea

Introducing the expectation values
\be N_q(\eta) = \langle b^\dagger_{\vq} b_{\vq} \rangle = \mathrm{Tr} \Big(b^\dagger_{\vq} b_{\vq}\,\rho_r(\eta) \Big) ~~;~~ M_q(\eta) = \langle b_{\vq} b_{-\vq} \rangle = \mathrm{Tr} \Big(b_{\vq} b_{-\vq}\,\rho_r(\eta) \Big) \label{exvals}\ee we find
\bea N'_q(\eta) & = &  \gamma_q(\eta)N_q(\eta) + i \Big[M_q(\eta)\beta_q(\eta)-M^*_q \beta^*_q(\eta) \Big] + S^{(N)}_q(\eta)\label{Ndot} \\
 M'_q(\eta) & = & i\alpha_q(\eta)M_q(\eta)+ 2i \beta^*_q(\eta) N_q(\eta)+ S^{(M)}_q(\eta)\,, \label{Mdot} \eea where
\bea \gamma_q(\eta) & = & \frac{4\lambda^2}{H^2\eta} \int^\eta_{\eta_0} \frac{d\eta'}{\eta'} K_I[q;\eta,\eta'] \, \textrm{Im}\Big[g(q,\eta')g^*(q,\eta) \Big] \label{gama}\\
S^{(N)}_q(\eta) & = & \frac{\lambda^2}{H^2\eta} \int^\eta_{\eta_0} \frac{d\eta'}{\eta'} \Big[ g(q,\eta)g^*(q,\eta')K[q;\eta,\eta']+g(q,\eta')g^*(q,\eta)K^*[q;\eta,\eta']\Big]\label{SN}\\
\beta_q(\eta) & = & \frac{2\lambda^2}{H^2\eta} \int^\eta_{\eta_0} \frac{d\eta'}{\eta'}K_I[q;\eta,\eta'] \,  \Big[g(q,\eta')g(q,\eta) \Big]\label{beta} \\
S^{(M)}_q(\eta) & = & -\frac{2\lambda^2}{H^2\eta} \int^\eta_{\eta_0} \frac{d\eta'}{\eta'}K[q;\eta,\eta'] g^*(q,\eta')g^*(q,\eta)\label{SM}\\
\alpha_q(\eta) & = & -\frac{2\lambda^2}{H^2\eta} \int^\eta_{\eta_0} \frac{d\eta'}{\eta'}K_I[q;\eta,\eta'] g(q,\eta')g^*(q,\eta)\label{alfa}\\
K[q;\eta,\eta'] & = & K_R[q;\eta,\eta']+i K_I[q;\eta,\eta'] \,.\label{KRKI} \eea

Once the kernel $K[q;\eta,\eta']$ is found the above equations can be integrated. The inhomogeneous source terms $S^{(N)}_q(\eta);S^{(M)}_q(\eta)$ are noteworthy, if these vanish, the vacuum $N_q=0, M_q=0$ would remain a fixed point of the dynamics, therefore these source terms indicate \emph{particle production} and the production of \emph{correlated pairs of particles}. Since these terms are independent of the initial number of particles and emerge even when the initial density matrix corresponds to the vacuum pure state, they can be understood perturbatively.

\section{Perturbative interpretation: particle production,   and entanglement}\label{sec:PI}

Before proceeding to analyze the full quantum master equation, it proves illuminating to understand the different contributions from the point of view of a perturbative expansion in the coupling. This analysis yields important insight into the correlations between the system and the enviroment and how the quantum master equation provides a non-perturbative resummation of self-energy contributions.

Consider that the initial density matrix describes the pure vacuum state $\rho_I(\eta_0) = |0\rangle \langle 0|$ where
$|0\rangle = |0\rangle_\chi\,|0\rangle_\psi$, up to second order in the interaction we find the state
\be |\Psi(\eta)\rangle = |0\rangle+ |\Psi^{(1)}(\eta)\rangle+ |\Psi^{(2)}(\eta)\rangle +\cdots \label{Psistate}\ee where
\bea && |\Psi^{(1)}(\eta)\rangle = (-i) \int^{\eta}_{\eta_0} H_I(\eta')\,d\eta'  \,|0\rangle \,,\label{Psi1}\\
&&|\Psi^{(2)}(\eta)\rangle = (-i)^2 \int^\eta_{\eta_0} H_I(\eta_1)\int^{\eta_1}_{\eta_0} H_I(\eta_2)\,d\eta_1  d\eta_2 \,|0\rangle  = (-i)  \int^\eta_{\eta_0} H_I(\eta_1)|\Psi^{(1)}(\eta_1)\rangle\,d\eta_1\,.\label{Psi2}\eea
Up to second order  the reduced density matrix up   is given by
\be \rho_r(\eta) = \mathrm{Tr}_{\psi} |\Psi(\eta)\rangle \langle\Psi(\eta)| =  \mathrm{Tr}_{\psi} \Bigg[|0\rangle \langle 0|+|\Psi^{(1)}(\eta)\rangle \langle \Psi^{(1)}(\eta)|+ |0\rangle \langle \Psi^{(2)}(\eta)|+|\Psi^{(2)}(\eta)\rangle \langle 0|\Bigg]\,. \label{rho2nd}\ee There is no cross term $|0\rangle \langle \Psi^{(1)}(\eta)|$ since the state $|\Psi^{(1)}(\eta)\rangle$ has vanishing overlap with $|0\rangle_\psi$  because of the normal ordering in the interaction Hamiltonian (see below).

\begin{figure}[ht!]
\begin{center}
\includegraphics[height=2.0in,width=2.0in,keepaspectratio=true]{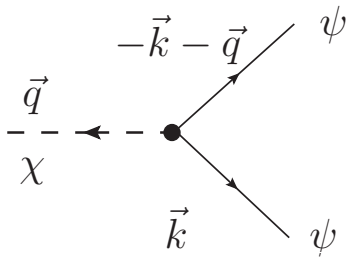}
\caption{The state $|\Psi^{(1)}(\eta)\rangle$: production of correlated $\chi$ and $\psi$ particles from the vacuum state. }
\label{fig:prod1}
\end{center}
\end{figure}

We find
\be |\Psi^{(1)}(\eta)\rangle = i\frac{\lambda}{H\sqrt{V}}\,\sum_{\vq}\sum_{\vk}\int^{\eta}_{\eta_0}\frac{d\eta_1}{\eta_1}\,g^*(q,\eta_1)u^*(k,\eta_1)
u^*(p,\eta_1) \,|1_{\vq}\rangle_\chi\,|1_{\vk},1_{\vp}\rangle_\psi~~;~~ \vp=-\vk-\vq \,,\label{psi1ofeta}\ee where
$|1_{\vq}\rangle_\chi$ are single $\chi$ particle states and similarly for the kets with $\psi$ particles. The state (\ref{psi1ofeta}) is an \emph{entangled quantum state} of the system ($\chi$)  and environmental ($\psi$) fields. This state describes particle production from the vacuum state by the interaction, depicted in fig.(\ref{fig:prod1}). In Minkowski space time these are virtual processes as they do not conserve energy, however in an expanding cosmology these processes are available as ``real'' because there is no time-like Killing vector\cite{woodardcosmo,decayds}. In particular the mode functions $g(q,\eta_1)$ feature a growing component when $q$ crosses the Hubble radius $g(q,\eta) \simeq \eta^{\frac{1}{2}-\nu_\chi}$ leading to \emph{secular} (growing) contributions to the time integrals as $\eta \rightarrow 0$ for $\nu_\chi \approx 3/2$ for a minimally coupled scalar field. This feature will become important below when we discuss the power spectrum.

\begin{figure}[ht!]
\begin{center}
\includegraphics[height=3.7in,width=3.7in,keepaspectratio=true]{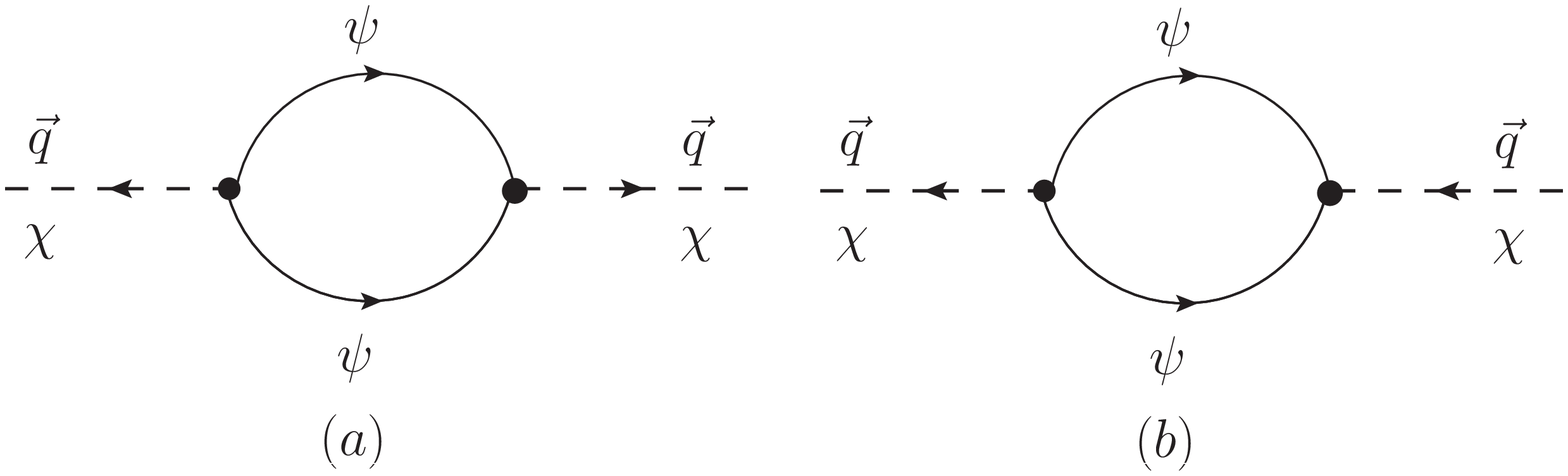}
\caption{The states in $|\Psi^{(2)}(\eta)\rangle$ that contribute to $\rho_r(\eta)$: $(a)= |\Psi^{(2)}(\eta)\rangle_{II} ~;~ (b) = |\Psi^{(2)}(\eta)\rangle_{I} $ . }
\label{fig:prod2}
\end{center}
\end{figure}

 In second order the state $|\Psi^{(2)}(\eta)\rangle$ features several contributions obtained by applying the interaction Hamiltonian to $|\Psi^{(1)}(\eta)\rangle$ as per the second equality in (\ref{Psi2}). However, only two of these contribute to $\rho_r(\eta)$ to second order, these are: I) annihilate \emph{all} particles in $|\Psi^{(1)}(\eta)\rangle$ returning to the full vacuum state $|0\rangle$, II) create another $\chi$ particle and annihilate (both) $\psi$ particles returning to the $\psi$ vacuum state but to a \emph{two particle state} of the $\chi$ field. Only these two contributions feature an overlap with the $\psi$ vacuum state necessary for the third and fourth terms in (\ref{rho2nd}). These two states are given respectively by
\be |\Psi^{(2)}(\eta)\rangle_I = -\frac{\lambda^2}{H^2} \int^{\eta}_{\eta_0}\frac{d\eta_2}{\eta_2}\int^{\eta_2}_{\eta_0}\frac{d\eta_1}{\eta_1} \sum_{\vq} g^*(q,\eta_1) g(q,\eta_2)\,K[q;\eta_2,\eta_1] |0\rangle_{\chi} |0\rangle_{\psi} \label{psi1I} \ee
 \be |\Psi^{(2)}(\eta)\rangle_{II} = -\frac{\lambda^2}{H^2} \int^{\eta}_{\eta_0}\frac{d\eta_2}{\eta_2}\int^{\eta_2}_{\eta_0}\frac{d\eta_1}{\eta_1} \sum_{\vq} g^*(q,\eta_1) g^*(q,\eta_2)\,K[q;\eta_2,\eta_1] |1_{\vq},1_{-\vq}\rangle_{\chi} |0\rangle_{\psi} \,.\label{psi1II} \ee and their associated Feynman diagrams are shown in fig.(\ref{fig:prod2})

The state $|\Psi^{(2)}(\eta)\rangle_{II}$ describes the production of correlated \emph{particle pairs}  of the field $\chi$ out of the initial vacuum state.
 Inserting these results into (\ref{rho2nd}) and carrying out the trace over the $\psi$ degrees of freedom, we find up to second order
\bea \rho_r(\eta) & = &  |0\rangle \langle 0|+ \sum_{\vq}\Bigg[ C^{(1)}(q;\eta)  |1_{\vq}\rangle  \langle 1_{\vq}| +  C^{(2)}_2(q,\eta)|1_{\vq},1_{-\vq}\rangle \langle 0|+  (C^{(2)}_2(q,\eta))^* | 0\rangle \langle 1_{-\vq},1_{\vq}\,| \nonumber \\ & + & C^{(2)}_0(q,\eta)|0\rangle \langle 0| + (C^{(2)}_0(q,\eta))^*|0\rangle \langle 0| \Bigg] \label{rho2nfin} \eea where now all the kets correspond to $\chi$ particle states (we suppressed the label $\chi$ to simplify notation). The coefficients are given by
\bea C^{(1)}(q;\eta) & = & \frac{\lambda^2}{H^2} \int^{\eta}_{\eta_0}\frac{d\eta_2}{\eta_2}\int^{\eta}_{\eta_0}\frac{d\eta_1}{\eta_1}g(q,\eta_2)g^*(q,\eta_1)
K[q;\eta_2,\eta_1] >0  \label{C1}\\
C^{(2)}_2(q,\eta)& = & -\frac{\lambda^2}{H^2} \int^{\eta}_{\eta_0}\frac{d\eta_2}{\eta_2}\int^{\eta_2}_{\eta_0}\frac{d\eta_1}{\eta_1}   g^*(q,\eta_1) g^*(q,\eta_2)\,K[q;\eta_2,\eta_1] \label{C22} \\
C^{(2)}_0(q,\eta) & = & -\frac{\lambda^2}{H^2} \int^{\eta}_{\eta_0}\frac{d\eta_2}{\eta_2}\int^{\eta_2}_{\eta_0}\frac{d\eta_1}{\eta_1}   g(q,\eta_2) g^*(q,\eta_1) \,K[q;\eta_2,\eta_1] \label{C20} \eea
The coefficient $C^{(1)}(q;\eta)$ can be written in a more illuminating manner by introducing $\Theta(\eta_2-\eta_1)+\Theta(\eta_1-\eta_2)=1$ in the integrals, in the term with $\Theta(\eta_1-\eta_2)$ relabel $\eta_1 \leftrightarrow \eta_2$ and use the property $K[q;\eta_1,\eta_2]=K[q;\eta_2,\eta_1]^*$ to find the relation
\be C^{(1)}(q;\eta) = -\Big(C^{(2)}_0(q,\eta)+ (C^{(2)}_0(q,\eta))^*  \Big)\,. \label{iden} \ee This identity confirms unitarity in the total time evolution, since
\be \mathrm{Tr}_{\chi}\rho_r(\eta) = 1 + \sum_{\vq}\Big[C^{(1)}(q;\eta)+C^{(2)}_0(q,\eta)+ (C^{(2)}_0(q,\eta))^* \Big] = 1 = \mathrm{Tr}\rho(\eta_0)\,. \label{unitarity} \ee

 Several aspects stem from the above results:
 \begin{itemize}
\item  \textbf{i):} with $b^\dagger_{\vq}\, b_{\vq} = N_q$ it follows that
\be \mathrm{Tr}N_q \,\rho_r(\eta) = \langle N_q \rangle (\eta) = C^{(1)}(q;\eta) \,,\label{aveN} \ee
 the relation (\ref{iden}) leads to
\be   N'_q  (\eta) = \frac{\lambda^2}{H^2\,\eta} \int^{\eta}_{\eta_0}\frac{d\eta_1}{\eta_1}   \Big[ g(q,\eta) g^*(q,\eta_1) \,K[q;\eta,\eta_1] + g^*(q,\eta) g(q,\eta_1) \,K^*[q;\eta,\eta_1]\Big]\,, \label{dN}\ee this is precisely the source term $S^{(N)}_q(\eta)$ (\ref{SN}) in the rate equation (\ref{Ndot}), and makes manifest the production of $\chi$ particles from the \emph{vacuum state}, namely the \emph{decay} of the vacuum during the time evolution. Whereas in Minkowski space time these are virtual processes leading to the wave function renormalization of the vacuum, the lack of a time-like Killing vector as a consequence of the cosmological expansion makes these processes to contribute in the long time limit\cite{woodardcosmo,woodard1}.

The super-Hubble limit of the mode functions (in the nearly massless case) $g(q,\eta) \propto 1/\eta$ (\ref{smaleta}) leads to secular growth of   particle production and vacuum decay in perturbation theory. These secular contributions are effectively re-summed by the full quantum master equation. This statement will become clear below.

 \item \textbf{ii):} The probability of the $\chi$ vacuum state in the reduced density matrix is
\be Z(\eta) = \langle 0|\rho_r(\eta)|0\rangle =  1-\sum_{\vq}   N_q(\eta)   \,, \label{Zofeta}\ee which is interpreted as the vacuum wave function renormalization, namely the probability of finding the ``bare'' vacuum state in the full state time evolved from the vacuum. This result is in agreement with that found in ref.\cite{richboy} within the Wigner-Weisskopf approximation, if the number of particles does not saturate in time, the Bunch-Davies vacuum is unstable towards particle production consistent with the results of refs.\cite{polyakov,akhmedov,picon}. The expectation values $M^*_q(\eta) = \langle b^\dagger_{\vq} b^\dagger_{-\vq}\rangle $ describe  the production of \emph{correlated pairs}.

\item \textbf{iii):} The contribution $|\Psi^{(1)}(\eta)\rangle$ given by (\ref{psi1ofeta}) and in the trace (\ref{rho2nd}) clearly shows that the degrees of freedom of the system $\chi$ are \emph{entangled} with those of the environment $\psi$. In particular for $q \ll -1/\eta$ and $k \gg -1/\eta$, there is quantum entanglement between \emph{super-Hubble} modes of the system and \emph{sub-Hubble} modes of the environment. This is in agreement with the super-sub-horizon entanglement discussed in ref. \cite{lello}.

\item \textbf{ iv):} Although the last two terms in (\ref{rho2nfin}) can be combined, they arise from $|\Psi^{(2)}(\eta)\rangle_I\langle 0|$ and $|0\rangle {}_{I}\langle\Psi^{(2)}(\eta)|$ respectively, then writing the various contributions in (\ref{rho2nfin}) in terms of $b^\dagger_{\vq}~,~b_{\vq}$ we find
 \bea && \rho_r(\eta)   =    \rho_r(\eta_0) + \sum_{\vq}\Bigg[ C^{(1)}(q;\eta)~  b^\dagger_{\vq}\,\rho_r(\eta_0)\,b_{\vq} +  C^{(2)}_2(q,\eta)\,b^\dagger_{\vq} b^\dagger_{-\vq}\,\rho_r(\eta_0) + \nonumber\\ &  &   (C^{(2)}_2(q,\eta))^* \,\rho_r(\eta_0)\,b_{\vq}\,b_{-\vq}     +   C^{(2)}_0(q,\eta)\,b_{\vq}\,b^\dagger_{\vq}\,\rho_r(\eta_0) + (C^{(2)}_0(q,\eta))^*  \,\rho_r(\eta_0)\,b^\dagger_{\vq}\,b_{\vq}\Bigg]\,. \label{rho2nfinbb} \eea Upon using the relation (\ref{iden}) and taking the derivative with respect to $\eta$ one recognizes by inspection the similar terms in (\ref{bbdagrho}) when taking $\rho_r(\eta) \rightarrow \rho_r(\eta_0) = |0\rangle \langle 0|$ on the right hand side of (\ref{bbdagrho})   to   lowest order in perturbation theory.

  \end{itemize}

  Therefore it becomes evident that (\ref{bbdagrho}) along with (\ref{Ndot},\ref{Mdot}) furnish a non-perturbative resummation for the time evolution, similar to quantum kinetic equations as discussed in ref.\cite{boyeff} in Minkowski space-time.

\section{Tracing out Sub-Hubble modes:}\label{sec:subhubble}
The discussion in the previous section has been quite general and does not specify in detail either the mass or the coupling to gravity of  the environmental degrees of freedom $\psi$,  namely the value of $\xi_\psi$ in (\ref{massds2}).

Our main goal in this article  is to understand the effective evolution of super-Hubble fluctuations of the system field $\chi$ (inflaton) upon integrating out (tracing over) fluctuations of the environmental field  with wavelengths that remain sub-Hubble all throughout inflation. In other words, we seek to obtain an effective field theory for long wavelength modes, those that would be of cosmological relevance today,  tracing out short wavelength modes of the environmental fields. For physical wavelengths that are much smaller than the Hubble radius $-k\eta \gg 1$, the Bunch-Davis mode functions of the environmental fields
\be u(k,\eta) ~~ \overrightarrow{-k\eta \gg 1} ~~ \frac{e^{-ik\eta}}{\sqrt{2k}}\,. \label{subH}\ee If the environmental fields are scalar fields minimally coupled to gravity, their quantum fluctuations feature a growing mode that is amplified upon becoming super-Hubble.    If the $\chi$ is the source of (adiabatic) perturbations a minimally coupled environmental scalar  field $\psi$ would yield isocurvature (or entropy) perturbations, which are   constrained by CMB observations. This motivates us to consider the bosonic field $\psi$ to be \emph{massless and conformally coupled to gravity}, namely $\xi_\psi = 1/6$ in which case $\nu_\psi =1/2$ and
\be u(k,\eta) = \frac{e^{-ik\eta}}{\sqrt{2k}} \label{conf}\ee for \emph{all} $k,\eta$. This is also the case if the bosonic field $\chi$ couples to a fermionic field with mass $m_f \ll H$, the mode functions for the fermionic fields are those of Minkowski space time but in terms of conformal time\cite{fermionswoodpro,uzan,boydVS}. Therefore, choosing $\psi$ to be  a massless conformally coupled scalar field is a \emph{proxy} for integrating out (tracing over) sub-Hubble degrees of freedom, leaving an effective action for the degrees of freedom that   become super-Hubble during inflation.
With the mode functions (\ref{conf}) it is now straightforward to obtain
\be K[q,\eta,\eta'] =   -\frac{i}{8\pi^2} \, \frac{e^{-iq(\eta-\eta')}}{(\eta-\eta'-i\varepsilon)}~~;~~ \varepsilon \rightarrow 0^+ \label{kernel} \ee where $\varepsilon$ is a short-distance cutoff that regulates the momentum integral in (\ref{kernels}). This result can also be confirmed from the operator product expansion of the composite operator $:\psi^2:$  in Minkowski space-time since the mode functions are the flat space time plane waves and the normal ordering is in the vacuum state. If, instead, we considered the environmental field ($\psi$) to be minimally coupled, there is an \emph{additional}  contribution to $K[q,\eta,\eta']$ from modes that become super-Hubble, this is discussed in the second reference in\cite{boyan} and will not be considered further here.

 Therefore $K[q,\eta,\eta']$ features a local and a non-local contribution,
\be  K[q,\eta,\eta'] =   -\frac{i}{8\pi^2}  {e^{-iq(\eta-\eta')}}\,\mathcal{P}\Big[\frac{1}{\eta-\eta'} \Big] + \, \frac{1}{8\pi}\,\delta(\eta-\eta')\,, \label{LNL}\ee where $\mathcal{P}$ stands for the principal part. The result (\ref{LNL}) is important, the correlation functions of environmental fields are often taken to be local in time, namely $\propto \delta(\eta-\eta')$ as in the second term in (\ref{LNL}), however even for fields with sub-Hubble fluctuations there is a non-local contribution which describes a long range memory. As it will be shown below, this non-local contribution is of paramount importance in the quantum master equation, it is also present if the environmental fields are minimally coupled to gravity (see second reference in\cite{boyan}).

It is convenient to write the non-local term in (\ref{LNL}) as
\be \mathcal{P}\Big[\frac{1}{\eta-\eta'} \Big] = \frac{\eta-\eta'}{(\eta-\eta')^2+\varepsilon^2} =
-\frac{1}{2}\, \frac{d}{d\eta'}\,\ln\Bigg[\frac{(\eta-\eta')^2+\varepsilon^2 }{(-\overline{\eta})^2}\Bigg]\,. \label{PP}\ee In (\ref{PP}) we have introduced an arbitrary scale $(-\overline{\eta})$ to render the argument of the logarithm dimensionless and acts as a subtraction or renormalization scale just as in the usual renormalization program in Minkowski space-time. It will be judiciously chosen below.  We can now input (\ref{LNL}) with (\ref{PP}) in (\ref{qme1}) and separate the local contribution from the $\delta(\eta-\eta')$ and the  non-local contribution from the principal part  to the quantum master equation (\ref{qme1}). The non-local contributions    feature the integrals
\be \int^{\eta}_{\eta_0} K[q;\eta,\eta'] \chi_{-\vq}(\eta')\,\frac{d\eta'}{\eta'}~~;~~\int^{\eta}_{\eta_0} K^*[q;\eta,\eta'] \chi_{-\vq}(\eta')\,\frac{d\eta'}{\eta'} \,,\label{nloc}\ee the form (\ref{PP}) allows to extract the divergence from  the non-local contribution upon integration by parts.

At this point we choose the arbitrary renormalization scale $\overline{\eta} = \eta_0$ with $-\eta_0$ very large (the beginning of the inflationary stage)  so that the wave vectors $q$ of interest are sub-Hubble at this time, namely $-q\eta_0 \gg 1$. When evaluated at $\eta_0$ the lower limit of the integrals (\ref{nloc}) , and for $-\eta \rightarrow 0$  the logarithm in (\ref{PP}) yields a contribution $\simeq \eta/\eta_0$ which, when combined with the  large $-q\eta$ behavior of $g(q,\eta)$ yield a contribution of order $\lambda^2/(H  q \eta_0)^2 \ll 1$ to the quantum master equation. Therefore,  choosing $\overline{\eta}  = \eta_0$ in (\ref{PP}) allows to neglect the contribution from lower limit in  the integration by parts. The manifestation of a renormalization group invariance of this choice will be discussed below.

We obtain, $\rho'_{r}(\eta) = \rho'_{rL}(\eta)+\rho'_{rNL}(\eta)$ where
\bea && \rho'_{rL}(\eta)   =   -i \frac{\lambda^2}{8\pi^2 H^2\eta^2}  \ln\Big[\frac{\varepsilon}{-\eta_0} \Big]~\sum_{\vq}\Big[\chi_{\vq}(\eta)\chi_{-\vq}(\eta),\rho_r(\eta) \ \Big] \nonumber \\ &-&
\frac{\lambda^2}{16\pi H^2 \eta^2} \sum_{\vq} \Bigg\{\chi_{\vq}(\eta)\chi_{-\vq}(\eta)\rho_r(\eta) + \rho_r(\eta)\chi_{\vq}(\eta)\chi_{-\vq}(\eta)-2\chi_{-\vq}(\eta)\rho_r(\eta)\chi_{\vq}(\eta) \Bigg\}\,, \label{rholoc}\eea for the local contribution, the non-local contribution is found to be
\bea
\rho'_{rNL}(\eta) & = & i\frac{\lambda^2}{8\pi^2 H^2\,\eta} \,\sum_{\vq} \Bigg\{ \chi_{\vq}(\eta)X_{-\vq}(\eta) \rho_r(\eta)\,     - \rho_r(\eta) \overline{X}_{-\vq}(\eta)  \chi_{\vq}(\eta)  \nonumber \\ &+&  \chi_{\vq}(\eta)\rho_r(\eta)\overline{X}_{-\vq}(\eta)\, -
X_{-\vq}(\eta) \rho_r(\eta) \chi_{\vq}(\eta)\Bigg\} \label{qmeNL}\eea where
\bea X_{-\vq}(\eta) & = & \int^{\eta}_{\eta_0} \frac{d}{d\eta'} \,\Bigg[e^{-iq(\eta-\eta')} \frac{\chi_{-\vq}(\eta')}{\eta'} \Bigg] \ln\Big[\frac{\eta-\eta'}{-\eta_0} \Big] d\eta'\label{Xdef} \\
\overline{X}_{-\vq}(\eta) & = & \int^{\eta}_{\eta_0} \frac{d}{d\eta'} \,\Bigg[e^{iq(\eta-\eta')} \frac{\chi_{-\vq}(\eta')}{\eta'} \Bigg] \ln\Big[\frac{\eta-\eta'}{-\eta_0} \Big]d\eta'\,.
  \label{Xovdef}\eea

The first term on the right hand side of the local contribution (\ref{rholoc}) (the commutator) is identified as an ultraviolet divergent mass renormalization, indeed this term can be written as
\be   \frac{-i\lambda^2}{8\pi^2 H^2\eta^2}  \ln\Big[\frac{\varepsilon}{-\eta_0} \Big]~\sum_{\vq}\Big[\chi_{\vq}(\eta)\chi_{-\vq}(\eta),\rho_r(\eta) \ \Big] = -i \Big[\delta H(\eta),\rho_r(\eta)\Big]   \,, \label{deltaH} \ee with
\be \delta H (\eta)= \frac{\delta M^2(\eta_0)}{2 H^2\eta^2} \,\sum_{\vq} \chi_{\vq}(\eta)\chi_{-\vq}(\eta) ~~;~~ \delta M^2(\eta_0) = \frac{\lambda^2}{4\pi^2 }  \ln\Big[\frac{\varepsilon}{-\eta_0} \Big] \,, \label{massren} \ee $\delta M^2(\eta_0)$ is an ultraviolet divergent mass renormalization of the same form as in Minkowski space time. This contribution is absorbed into a mass renormalization, the renormalized mass is
\be M^2_R(\eta_0) = M^2_{\chi}+\delta M^2(\eta_0)  \label{Mren} \,,\ee
where we emphasized that this renormalized mass depends on the renormalization scale $\eta_0$. In what follows we  absorb this mass renormalization in the original Lagrangian and include a counterterm to precisely cancel  (\ref{deltaH})  thereby neglecting the first term in (\ref{rholoc}).

We can now insert the kernel (\ref{kernel}) into the coefficient functions (\ref{gama}-\ref{KRKI}) in the equations for   $N_q(\eta)$ and  $M_q(\eta)$ (\ref{Ndot}, \ref{Mdot}) and solve them to obtain their time evolution. However, while understanding their dynamical evolution merits such study on its own,  neither $N_q(\eta)$ nor $M_q(\eta)$ are directly observable. Instead an important observable is the power spectrum
\be \mathcal{P}(q;\eta) = \frac{q^3}{2\pi^2} \langle \phi_{\vq}(\eta) \,\phi_{-\vq}(\eta) \rangle =
 \frac{q^3\,H^2\,\eta^2}{2\pi^2} \langle \chi_{\vq}(\eta) \,\chi_{-\vq}(\eta) \rangle \,. \label{pofqpo}\ee  In terms of the field expansion and the expectation values (\ref{exvals}) it is given by
 \be \mathcal{P}(q;\eta) =
 \frac{q^3\,H^2\,\eta^2}{2\pi^2} \Bigg[(1+2N_q(\eta))\,|g(q,\eta)|^2 + M_q(\eta) \, g^2(q,\eta)+  M^*_q(\eta) \, (g^*(q,\eta))^2 \Bigg]\,. \label{pofqpo2}\ee  Furthermore, only the power spectrum for wavevectors of cosmological relevance are of observational interest, these correspond to wavelengths that became larger than the Hubble radius   $\simeq 10$ e-folds before the end of inflation. The quantum fluctuations corresponding to these mode functions are amplified upon becoming super-Hubble and become classical, the mode functions feature a growing and a decaying mode and only the growing mode is relevant when the fluctuations become super-Hubble. This means that only a particular combination of $N_q(\eta)$ and $M_q(\eta)$ determined by the growing modes is relevant to the power spectrum. Rather than solving the coupled set of equations (\ref{Ndot},\ref{Mdot}) it is more convenient to directly study the dynamical evolution of the relevant combinations. Therefore we re-cast the quantum master equation in terms of the degrees of freedom that have direct relevance to the    power spectrum of super-Hubble   fluctuations.

\section{Classicalization and correction to the power spectrum:}
Rather than studying the dynamics of particle production and two-particle correlations and solving the coupled equations (\ref{Ndot}, \ref{Mdot}) with the time dependent coefficients determined by the kernel $K[q,\eta,\eta']$ above,  we focus on understanding the impact of   tracing of sub-Hubble modes upon physical observables, in particular the power spectrum.

 The modes of a   bosonic field minimally coupled to gravity become classical when their physical wavelength becomes larger than the Hubble radius during an inflationary stage (superhorizon), this is a consequence of the fact that there is a growing mode that becomes amplified and a decaying mode whose amplitude diminishes\cite{polarski}. This is manifest in the mode functions $g(q,\eta)$ (\ref{gqeta}) since
 \be H^{(1)}_{\nu}(-q\eta) = J_{\nu}(-q\eta)+i \, Y_{\nu}(-q\eta) \label{Hmode} \ee and
 \be J_\nu(z) ~~\overrightarrow{z\rightarrow 0} ~~\Big(\frac{z}{2}\Big)^{\nu} \frac{1}{\Gamma(\nu+1)}~~;~~ Y_\nu(z) ~~ \overrightarrow{z\rightarrow 0} ~~ -\Big(\frac{z}{2}\Big)^{-\nu}\, \frac{\Gamma(\nu)}{\pi} \,. \label{growdec}\ee In order to exhibit the classicalization of the quantum fluctuations in a more direct manner, it proves convenient to use the real mode functions $J_\nu;Y_\nu$ corresponding to the decaying and growing modes respectively to expand the field, this is achieved by introducing   the   combinations
 \bea Q_{\vq} & =  & \frac{1}{\sqrt{2}} \,\Big( b_{\vq}\,e^{i\frac{\pi}{2}(\nu_\chi+\frac{3}{2})}+ b^\dagger_{-\vq}\,e^{-i\frac{\pi}{2}(\nu_\chi+\frac{3}{2})} \Big) ~~;~~ Q^\dagger_{\vq} = Q_{-\vq} \label{Qs} \\  P_{\vq}
&  = &  \frac{i}{\sqrt{2}} \,\Big( b^\dagger_{-\vq}\,e^{-i\frac{\pi}{2}(\nu_\chi+\frac{3}{2})}  - b_{\vq}\,e^{i\frac{\pi}{2}(\nu_\chi+\frac{3}{2})}\Big)~~;~~ P^\dagger_{\vq} = P_{-\vq}\,. \label{Ps} \eea
 These $P's$ and $Q's$ are canonical variables obeying the canonical commutation relations
\be [P^\dagger_{\vq},Q_{\vk}] = -i \delta_{\vq,\vk}~~;~~[P_{\vq},P_{\vk}]=[Q_{\vq},Q_{\vk}]=0 \,. \label{CCR} \ee Introducing the \emph{real} growing and decaying mode functions
\be g_+(q;\eta) = \sqrt{\frac{-\pi\eta}{2}}~ Y_{\nu_\chi}(-q\eta) ~~;~~ g_-(q;\eta) = \sqrt{\frac{-\pi\eta}{2}}~ J_{\nu_\chi}(-q\eta)\,, \label{gpms}\ee we can now write the field expansion (\ref{chiofq}) as
\be \chi_{\vq}(\eta) = Q_{\vq}\,  g_+(q;\eta)+ P_{\vq}\,   g_-(q;\eta) \,.\label{chipq} \ee
As discussed in refs.\cite{polarski} the classicalization of fluctuations in the super-Hubble limit is gleaned from the relation between $\chi_{\vq}$ and its conjugate momentum
\be \Pi_{\vq}(\eta) = \chi'_{\vq}(\eta) = Q_{\vq}\,g'_+(q;\eta)+ P_{\vq}\,g'_-(q;\eta)~~;~~[\Pi_{-\vk}(\eta),\chi_{\vq}(\eta)] = -i \delta_{\vk,\vq} \,.  \label{Pichi}\ee
For $-q\eta \rightarrow 0$ the growing solution $g_+(q;\eta)$ dominates and
\be \Pi_{\vq}(\eta)\simeq \chi_{\vq}(\eta)\, \Big(\frac{g'_+(q;\eta)}{g_+(q;\eta)} \Big) \propto \frac{\chi_{\vq}(\eta)}{\eta}\,,\label{Pipropchi} \ee namely the commutator between $\Pi,\chi$  becomes much smaller than the amplitude of the canonical variables, a necessary condition for  the classicalization of the fields. This argument is independent of the interaction with other fields in the theory and relies solely on the fact that the   solutions of the Heisenberg equations of motion for a \emph{minimally coupled scalar field} feature a growing and a decaying mode in the long time limit after the particular physical wavelength has become super-Hubble. This feature does not apply to either fermionic fields (which are never classical) nor to massless conformally coupled scalar fields (at least in absence of interactions).

The power spectrum of the original (un-scaled) field $\phi$ is
\be \mathcal{P}(q;\eta) = \frac{q^3}{2\pi^2}\,\langle \phi_{\vq}(\eta) \phi_{-\vq}(\eta)\rangle \label{pofq}\ee  in the limit $-q\eta \ll 1$, where $\phi_{\vq}$ is the spatial Fourier transform of the original field. From the scaling relation (\ref{rescale}) and (\ref{aofeta}) and with $\chi_{\vq}$ expanded as in (\ref{chipq}) we find for $-q\eta \rightarrow 0$
\bea \mathcal{P}(q;\eta) & = &  \frac{H^2\,2^{2\nu_\chi}\,\Gamma^2(\nu_{\chi})}{4\pi^3}\big( -q\eta \big)^{3-2\nu_{\chi}} \Bigg[\langle Q^\dagger_{\vq} Q_{\vq} \rangle - \frac{\pi}{2^{2\nu_{\chi}}\,\nu_{\chi}\Gamma^2(\nu_{\chi})}\, \langle \Big(Q_{\vq}P^\dagger_{\vq} +P_{\vq} Q^\dagger_{\vq}\Big)\rangle \,\big( -q\eta\big)^{2\nu_\chi} \nonumber \\ & + &  \Big(\frac{\pi}{2^{2\nu_{\chi}}\,\nu_{\chi}\Gamma^2(\nu_{\chi})} \Big)^2\, \langle P^\dagger_{\vq}P_{\vq} \rangle\, \big(-q\eta\big)^{4\nu_\chi} \Bigg] \label{pow}\eea
With  $\phi$ being a minimally coupled scalar field with $M_{R} \ll H$, where now $M_R$ is the renormalized mass (\ref{Mren}) (here we suppressed the renormalization scale $\eta_0$ in the definition of the renormalized mass) it follows that
\be \nu_{\chi} \simeq \frac{3}{2} - \frac{M^2_R}{3H^2} \,, \label{nuchi}\ee therefore to leading order in $M_{R}/H$ we find
\be  \mathcal{P}(q;\eta) \simeq \frac{H^2}{2\pi^2} \big( -q\eta \big)^{\frac{2M^2_R}{3H^2}}\, \Bigg[ \langle Q^\dagger_{\vq} Q_{\vq} \rangle - \frac{1}{3}\, \langle \Big(Q_{\vq}P^\dagger_{\vq} +P_{\vq} Q^\dagger_{\vq}\Big)\rangle \,\big( -q\eta \big)^{3}   +    \frac{1}{9}\, \langle P^\dagger_{\vq}P_{\vq} \rangle\, \big(-q\eta\big)^{6}  \Bigg]\,. \label{powMH} \ee A few e-folds after the corresponding mode crosses the Hubble radius, namely $-q\eta \ll 1 $ the second and third terms in the bracket can be safely neglected and
\be \mathcal{P}(q;\eta) \,\, \, \overrightarrow{-q\eta \rightarrow 0 } \,\, \,\frac{H^2}{2\pi^2} \big( -q\eta \big)^{\frac{2M^2_R}{3H^2}}\,   \langle Q^\dagger_{\vq} Q_{\vq} \rangle \label{superP} \ee When the state is the Bunch-Davies vacuum $\langle Q^\dagger_{\vq} Q_{\vq} \rangle=1/2$ and for $M_R \rightarrow 0$ one finds the usual  scale invariant power spectrum for a massless scalar field,
\be \mathcal{P}(q;\eta) \,\, \, \overrightarrow{-q\eta \rightarrow 0 } \,\, \,\frac{H^2}{(2\pi)^2} \,. \label{Pscaleinv}\ee In order to obtain the quantum corrections from tracing out the sub-Hubble degrees of freedom, the average $\langle (\cdots)\rangle$ must be carried out with the reduced density matrix $\rho_r(\eta)$. The free-field Heisenberg operator (\ref{chipq}) is precisely the operator in the interaction picture wherein $Q_{\vq},P_{\vq}$ do \emph{not} depend on time, then the average in the bracket in (\ref{powMH}) is now understood with $\rho_r(\eta)$ and in the limit $(-q\eta) \rightarrow 0$ the terms with $(-q\eta)^3,(-q\eta)^6$ can be safely neglected, leaving only the contribution $\langle Q^\dagger_{\vq} Q_{\vq} \rangle$. Namely  we  need to obtain
\be \mathcal{P}(q;\eta) =\frac{H^2}{2\pi^2} \, \big( -q\eta \big)^{\frac{2M^2_R}{3H^2}}\,\,   \mathrm{Tr}\Big( Q^\dagger_{\vq} Q_{\vq} \, \rho_r(\eta)\Big) \,.\label{superPrhor} \ee Therefore it is more convenient to write $\rho_r(\eta)$ in terms of the canonical phase space variables $Q_{\vq},P_{\vq}$ instead of the creation and annihilation operators. Furthermore, we need to extract the coefficient functions of $Q_{\vq}, P_{\vq}$ in the expressions $X_{-\vq},  \overline{X}_{-\vq}$ in (\ref{Xdef},\ref{Xovdef}). Since the power spectrum is obtained in the super-Hubble limit, namely $-q\eta \rightarrow 0$ we can simplify the integrals in (\ref{Xdef},\ref{Xovdef}) by i) taking $q\eta \rightarrow 0$, ii) cutting off- the integrals at a time scale $\eta_* \simeq -1/q$ in the lower limits of the integrals, since the integrand is dominated by the late time $\eta' \simeq \eta  \rightarrow 0$ region. These approximations capture the leading behavior in the super-Hubble limit and simplify the integrals in (\ref{Xdef},\ref{Xovdef})) leading to
\be  \overline{X}_{-\vq} = X_{-\vq}(\eta)   =   Q_{-\vq} \, \overline{g}_+(q;\eta)+   P_{-\vq} \, \overline{g}_-(q;\eta) \,, \label{capXs} \ee where
\be \overline{g}_{\pm}(q;\eta) =  \int^{\eta}_{\eta_*} \frac{d}{d\eta'} \,\Bigg[  \frac{g_{\pm}(q;\eta')}{\eta'} \Bigg] \ln\Big[\frac{\eta-\eta'}{-\eta_0} \Big] d\eta' ~~;~~ -q \eta_* \simeq 1 \,. \label{gpmdef} \ee
After renormalization of the mass by a mass counterterm to cancel the first term in $\rho'_{rL}(\eta)$ (\ref{rholoc}) we find the local part of the quantum master equation to be
\bea \rho'_{rL}(\eta) & = & -\frac{\lambda^2}{16\pi H^2\eta^2} \sum_{\vq} \Bigg\{ \Big[ Q_{\vq}  Q_{-\vq}\,\rho_r(\eta)+ \rho_r(\eta)Q_{\vq}  Q_{-\vq}-2 Q_{-\vq}\rho_r(\eta) Q_{\vq}  \Big]\,\big(g_+(q;\eta)\big)^2 \nonumber \\ & + &  \Big[ P_{\vq}  P_{-\vq}\,\rho_r(\eta)+ \rho_r(\eta)P_{\vq}  P_{-\vq}-2 P_{-\vq}\rho_r(\eta) P_{\vq}  \Big]\,\big(g_-(q;\eta)\big)^2 \nonumber \\ & + & \Big[ Q_{\vq}  P_{-\vq}\,\rho_r(\eta)+ \rho_r(\eta)Q_{\vq}  P_{-\vq}-2 P_{-\vq}\rho_r(\eta) Q_{\vq}  \Big]\,\big(g_+(q;\eta)g_-(q;\eta)\big) \nonumber \\ & + &
 \Big[ P_{\vq}  Q_{-\vq}\,\rho_r(\eta)+ \rho_r(\eta)P_{\vq}  Q_{-\vq}-2 Q_{-\vq}\rho_r(\eta) P_{\vq}  \Big]\,\big(g_+(q;\eta)g_-(q;\eta)\big) \Bigg\} \,,\label{rhoLQP}\eea and the non-local contribution is
 \bea  \rho'_{rNL}(\eta)   & = &   \frac{i\lambda^2}{8\pi^2 H^2 \eta} \sum_{\vq} \Bigg\{g_+(q;\eta) \overline{g}_+(q;\eta)\,\Big[ Q_{\vq}  Q_{-\vq}\,\rho_r(\eta)- \rho_r(\eta)Q_{\vq}  Q_{-\vq} \Big] \nonumber \\ & + &  g_-(q;\eta) \overline{g}_-(q;\eta)\,\Big[ P_{\vq}  P_{-\vq}\,\rho_r(\eta)- \rho_r(\eta)P_{\vq}  P_{-\vq} \Big] \nonumber \\ & + &
  g_+(q;\eta) \overline{g}_-(q;\eta)\,\Big[Q_{\vq}P_{-\vq}\, \rho_r(\eta) - P_{-\vq} \rho_r(\eta)Q_{\vq} \Big] \nonumber \\ &-& g_+(q;\eta) \overline{g}_-(q;\eta)\,\Big[ \rho_r(\eta)\,P_{-\vq} Q_{\vq} -Q_{\vq} \rho_r(\eta) P_{-\vq} \Big] \nonumber \\ & + & \overline{g}_+(q;\eta) g_-(q;\eta)\Big[ P_{\vq} Q_{-\vq} \rho_r(\eta) - Q_{-\vq} \rho_r(\eta) P_{\vq} \Big] \nonumber \\ & - &
\overline{g}_+(q;\eta) g_-(q;\eta)\Big[ \rho_r(\eta) Q_{-\vq}P_{\vq}- P_{\vq} \rho_r(\eta)  Q_{-\vq}\Big]\Bigg\}\,. \label{rhoNLQP}\eea
As mentioned above, in order to obtain   $\langle Q_{\vq} Q_{-\vq}\rangle = \mathrm{Tr}Q_{\vq} Q_{-\vq}\,\rho_r(\eta) $ we would need to solve the quantum master equation. However, since in the interaction picture $Q_{\vq}$ is time independent, we obtain the equation of motion for $\langle Q_{\vq} Q_{-\vq}\rangle$ instead, namely
\be \frac{d}{d\eta}\langle Q_{\vq} Q_{-\vq}\rangle =  \mathrm{Tr}\Big( Q_{\vq} Q_{-\vq}\,\rho'_r(\eta)\Big)\,,  \label{eomQ2} \ee from which we obtain $\langle Q_{\vq} Q_{-\vq}\rangle$ by integration.

The terms bilinear in $P$    are proportional to $(g_-(q;\eta))^2/\eta^2 \propto \eta^2$ in $\rho_{rL}$ and $ g_-(q;\eta)/\eta \propto \eta$  in $\rho_{rNL}$ both are subleading in the long time limit $\eta \rightarrow 0$ and can be neglected in the evaluation of  $\langle Q_{\vq} Q_{-\vq}\rangle$. We find that $\rho'_{rL} $ yields no contribution to (\ref{eomQ2})  in the long time limit, finally we obtain the remarkable result
\be \frac{d}{d\eta}\langle Q_{\vq} Q_{-\vq}\rangle = - \Gamma(q;\eta)\,\langle Q_{\vq} Q_{-\vq}\rangle \,,\label{rateQ2}\ee  where
\be \Gamma(q;\eta) = \frac{\lambda^2\,\overline{g}_+(q;\eta) g_-(q;\eta)}{2\pi^2 H^2 \eta}\,  \label{gamma} \ee is determined by the \emph{non-local} contribution to the quantum master equation. The above equation is valid at long time so that the particular   wavevector $q$ has crossed the Hubble radius since we have neglected the contribution from the decaying mode. Therefore the initial condition for the integration of (\ref{rateQ2}) must be set at the scale $\eta_* \simeq -1/q$, hence
 \be \langle Q_{\vq} Q_{-\vq}\rangle (\eta) = e^{-\int^\eta_{\eta_*}\Gamma(q;\eta')d\eta'}\,\,\langle Q_{\vq} Q_{-\vq}\rangle (\eta_*) \,. \label{intQ2} \ee

We can extract the leading behavior from the integral in (\ref{gpmdef}) in the limit $-q\eta \rightarrow 0$, and for $M^2_R/H^2 \ll 1$ setting $\nu_\chi = 3/2$, namely
\be g_+(q;\eta) = \frac{1}{q^{3/2}\,\eta} ~~;~~ g_-(q;\eta) = \frac{1}{3}~ q^{3/2}\,\eta^2 \,,\label{gis}\ee from which we find to leading and next to leading order as $\eta \rightarrow 0$
\be \overline{g}_+(q;\eta) = \frac{1}{q^{3/2}\,\eta^2}\Big[ \ln \Big(\frac{\eta}{\eta_0} \Big) -1 \Big]\,. \label{gbar}\ee These results yield to leading and next to leading order as $\eta \rightarrow 0$ and $M^2_R/H^2 \ll 1$
\be \Gamma(q;\eta) =   \frac{\lambda^2}{6\pi^2 H^2 \eta}\,\Big[ \ln \Big(\frac{\eta}{\eta_0} \Big) -1 \Big] \,.\label{gammafin}\ee Carrying out the integral in (\ref{intQ2}) and inserting the result into the power spectrum (\ref{superPrhor}) we find
\be \mathcal{P}(q;\eta) = \frac{H^2}{2\pi^2} \,  \, e^{\alpha(q)\ln[-q\eta]}\,\,\, e^{-\frac{\lambda^2}{12\pi^2H^2}\ln^2[-q\eta]} \,\,\langle Q_{\vq} Q_{-\vq}\rangle (\eta_*) \,,\label{Pofqfina}\ee where
\be \alpha(q) =  \frac{2M^2_R(\eta_0)}{3H^2} + \frac{\lambda^2}{6\pi^2H^2}\Big[\ln[-q\eta_0]+1 \Big] \,. \label{alfaq}\ee
The result (\ref{Pofqfina}) is noteworthy: the quantum master equation yields a non-perturbative resummation of secular Sudakov-type double logarithms in the long time limit.

With $M^2_R(\eta_0)$ given by (\ref{Mren}) and $\delta M^2(\eta_0)$ by (\ref{massren}) it is clear that $\alpha(q)$ is \emph{independent} of the renormalization scale $\eta_0$,
 this is a reassuring confirmation of the consistency of the renormalization procedure: the effective renormalized mass changes with the renormalization scale, but the physical power spectrum is independent of this scale, this is a manifestation of a renormalization group invariance. While the q-independent term in the bracket in (\ref{alfaq}) can be absorbed as a finite renormalization of the mass the q-dependence of $\alpha(q)$ is a consequence of the logarithmic renormalization. It is clear that even when the non-interacting theory is massless and the power spectrum is  scale invariant, the coupling to the subhorizon degrees of freedom will induce a mass via renormalization effects, the logarithmic ultraviolet divergence implies that the renormalized mass depends on an arbitrary renormalization scale and in turn this results on a $q$ dependence of $\alpha(q)$.

 It remains to estimate $\langle Q_{\vq} Q_{-\vq}\rangle (\eta_*)$, to achieve this we would have to keep the full expressions for the mode  functions $g_\pm(q;\eta')$ inside the integrals in (\ref{Xdef},\ref{Xovdef}) along with the oscillatory factors $e^{iq\eta'}$. However for $-q\eta' \gtrsim 1$ in the integration the mode functions behave as $\cos[-q\eta']/\sqrt{q}; \sin[-q\eta']/\sqrt{q}$ and the integrands are of $\mathcal{O}(1)$  in the region of integration  $ -\eta_0 \gg -\eta' \gg -\eta_* = 1/q$. The mode functions amplify sharply for $-q\eta' < 1$ and the contribution from this time region dominates  the integrals. Therefore the region of integration with $ -\eta' > -\eta_*$ yields perturbatively small contributions of order $\lambda^2/H^2 \ll 1$, consequently $\langle Q_{\vq} Q_{-\vq}\rangle (\eta_*) = \langle Q_{\vq} Q_{-\vq}\rangle (\eta_0)+ \mathcal{O}(\lambda^2/H^2) \simeq 1/2+\mathcal{O}(\lambda^2/H^2)$ where we have used that at $\eta_0$ the initial density matrix describes the Bunch-Davies vacuum. Therefore the final result for the power spectrum for wavevectors that cross the Hubble radius is
 \be
\mathcal{P}(q;\eta) \simeq \frac{H^2}{(2\pi)^2}\,\,  e^{\alpha(q)\ln[-q\eta]}\,\,\, e^{-\frac{\lambda^2}{12\pi^2H^2}\ln^2[-q\eta]}\,. \label{powerfinal} \ee   Including the second term in the bracket in (\ref{alfaq}) into a finite renormalization of $M_R$   and setting the renormalized mass $M^2_R(\eta_0)=0$ so that in absence of coupling the power spectrum is scale invariant, namely
\be \alpha(q) =   \frac{\lambda^2}{6\pi^2H^2}\,\ln[-q\eta_0]  \,,\label{alfaqMzero}\ee
we obtain the final form of the power spectrum
 \be
\mathcal{P}(q;\eta) \simeq \frac{H^2}{(2\pi)^2}\,\,  e^{\frac{\lambda^2}{6\pi^2H^2}\Big[\ln[-q\eta_0] \ln[-q\eta] - \frac{1}{2}\,\ln^2[-q\eta]\Big]}\,. \label{powerfinal2} \ee
It follows from this result  that the power spectrum \emph{decays} when the wavevector $q$   becomes ``superhorizon'' $-q\eta < 1$. If the wavevector $q$ is of cosmological relevance today, it crossed the Hubble radius $\simeq 10$ e-folds before $\eta_f$, the end of inflation,  and if inflation lasts $\simeq 60 $ e-folds, it follows that
\be \ln[-q\eta_0] \simeq 50 ~;~ |\ln[-q\eta_f]| \simeq 10 \label{efs}\ee and the exponent in (\ref{powerfinal2}) $\simeq 8 \lambda^2/H^2 $.  With $\lambda^2/H^2 \ll 1$ under the assumption of weak coupling  and the realm of validity of the quantum master equation, we find that the decay of the power spectrum could be \emph{marginally} observable if $\lambda / H \simeq 0.1$ implying a suppression $\lesssim 10\%$ of the power spectrum for modes that re-enter the horizon near recombination, assuming no further corrections during the post-inflationary era.

Therefore, although the power spectrum decays as a consequence of the interaction with the environmental degrees of freedom, it is likely that these corrections are of marginal observational  relevance, at least within the model studied here. However, this important observational fact notwithstanding, there is the noteworthy and fundamental  aspect that the amplitude of the perturbation does not freeze out   but decays after   crossing the Hubble radius. These results confirm in a non-perturbative manner previous perturbative analysis\cite{sloth1,kahya} but also points out that not only the power spectrum does not freeze-out after ``horizon crossing'' but that the time dependence is associated with a violation of scale invariance even when in absence of interactions the power spectrum is exactly scale invariant.

\section{Discussion}\label{sec:discussion}
There are several aspects of the method and the results that merit   discussion:

\begin{itemize}

\item \textbf{What is being re-summed?:} The quantum master equation provides a non-perturbative resummation as is explicit in the final result for the power spectrum (\ref{powerfinal}), the question is what type of contributions are being re-summed. Tracing out the sub-Hubble degrees of freedom is manifest in the correlation function (\ref{ggreat},\ref{gless}) of the field $\psi$ shown in fig. (\ref{fig:correlators}). These correlation functions define the one-loop self-energy corrections to the field $\chi$. The perturbative analysis of section (\ref{sec:PI})   shows how these correlation functions enter in the time evolution of the density matrix and lead to the production of single particles and correlated pairs and the quantum entanglement between the quanta of the system ($\chi$) and those of the (traced-over) environment ($\psi$). The Sudakov-type double logarithms originate from several different contributions: i) the logarithmic behavior of the one-loop self-energy is the same as in Minkowski space time because these arise from conformally coupled massless fields that act as  \emph{proxies} for degrees of freedom that remain sub-Hubble all throughout inflation. ii) the scale factor $\propto 1/\eta$ in the interaction vertex, iii) the growing mode functions inside the kernel which lead to the factor $1/\eta^2$ in (\ref{gbar}), which is canceled by the decaying mode (see eqns. (\ref{gamma}, \ref{gis},\ref{gammafin})), thus leaving finally the $1/\eta$ from the scale factor at the vertex which enhances the long time limit logarithmically (proportional to the number of e-folds). Thus, in summary,\textbf{ the quantum master equation furnishes a non-perturbative resummation of the secular Sudakov-type double logarithms of the one-loop self-energy of the $\chi$ field}. When the mode functions of the $\chi$ field become ``super-horizon'' the quantum master equation effectively describes   quantum entanglement between these super-horizon modes and the sub-horizon degrees of freedom that have been traced over, in agreement with the results of ref.\cite{lello}.

\item \textbf{Generality:} We have considered the interaction $\propto \chi \psi^2$ where $\chi$ describe the degrees of freedom that are of cosmological relevance and become ``super-horizon'', namely the ``system'' and $\psi$ describes the sub-horizon degrees of freedom that are integrated out and is considered as a ``bath'' or environment. This interaction is, in fact, more general and describes various relevant cases: for example consider the case in which the original (unscaled) scalar field $\phi$ features a quartic self-interaction $g\,\phi^4$ during slow roll inflation when $\phi$ develops a nearly constant expectation value $\Phi$ (in slow roll). Writing $\phi(\vx,\eta) = \Phi + \delta\phi(\vx,\eta)$ (here we assumed slow roll and neglected the time dependence of $\Phi$) and writing the spatial Fourier transform of $\delta \phi(\vx,\eta)$ as $\delta \phi_>(\vq,\eta)  + \delta\phi_<(\vq,\eta) $  where $ \delta\phi_<(\vq,\eta) $ describes modes that are deep inside the Hubble radius all throughout inflation  in terms of the mode functions (\ref{conf}),  one obtains the vertex $(g\Phi) \,\delta \phi_>(\vq,\eta)\delta \phi_<(\vk,\eta)\, \delta \phi_<(-\vk-\vq,\eta)$ which is the of form $\lambda \chi \psi^2$ where $\chi = a(\eta) \delta\phi_>$, $\psi = a(\eta) \delta \phi_<$. In this interpretation tracing over the $\delta \phi_<$ degrees of freedom to obtain the reduced density matrix is akin to a \emph{Wilsonian} coarse graining procedure of integrating out short wavelength fluctuations leading to an effective field theory for long-wavelength fluctuations\cite{bala}. A Yukawa interaction with fermionic fields is also described by the cubic vertex considered here. Indeed the fermionic fields are expanded in mode functions of the form (\ref{uketa}) with the index $\nu \simeq 1/2 $ for a massive field with $m_f \ll H$ with technical differences associated with the spinorial contributions to the loop corrections, which however yield a similar logarithmic contribution to the self-energy after a quadratic renormalization of the mass since the loop correction is similar to that in Minkowski space time.

    \item \textbf{Minimally coupled scalars secular and infrared enhancements:} If the inflaton (``system'') scalar field couples to other (environmental) scalar fields that are minimally coupled to gravity and with masses $M\ll H$ these ``environmental'' fields feature quantum fluctuations that are infrared and secularly enhanced when their wavelengths become super-Hubble\cite{woodardcosmo,decayds,akhmedov,prokowood,sloth1,riotto,boyan,lello}.


        Although in this article we focused on influence of sub-Hubble correlations, the importance of the infrared and secular enhancements when the environmental scalar field is minimally coupled to gravity merits a discussion.

        To begin with, consider the case when the environmental field is \emph{minimally coupled and massless} in this case $\nu_\psi = 3/2$ and
        \be  u(k,\eta) = \frac{e^{-ik\eta}}{\sqrt{2k}} \,\big[1+ \frac{i}{k\eta} \big] \label{umm} \,,  \ee then it is clear that the correlation function  $K[q,\eta,\eta']$ in (\ref{kernels})  features logarithmic \emph{infrared divergences} in the integration regions $k \simeq 0 ~;~ |\vk+\vq| \simeq 0$. A non-vanishing mass for the $\psi$ field regulates the infrared, consider $M_\psi \ll H$ from which it follows that for a minimally coupled environmental field
        \be \nu_\psi \simeq \frac{3}{2} - \Delta  ~~;~~ \Delta  = \frac{M^2_\psi}{3H^2} \label{delsi}\,. \ee In ref.\cite{boyan} it is shown that the logarithmic infrared divergences of the (self-energy) kernel are manifest as poles in $\Delta$. The calculation of the kernel $K[q,\eta,\eta']$ follows the same steps described in detail in ref. \cite{boyan} (see the appendix in the second reference). The regions of integration $k \simeq 0~;~|\vk+\vq| \simeq 0$ are isolated and an infrared cutoff $\mu$ is introduced in these regions, within which the  small argument expansion $u(p,\eta) \propto p^{-\nu_\psi}$ is used and the integration yields poles in $\Delta$. Outside these infrared regions it is safe to take $\nu_\psi = 3/2$, the details are available in the second reference in \cite{boyan}. Using the results of this reference we find to leading and next to leading  order in $\Delta$
        \be K[q,\eta,\eta']   \equiv   K_{IR}[q,\eta,\eta']+K_{cc}[q,\eta,\eta']\,, \label{totKe}\ee where 
        \bea   K_{IR}[q,\eta,\eta'] & = &   \frac{H^{(1)}_{\nu_\psi}(-q\eta)\,H^{(2)}_{\nu_\psi}(-q\eta')}{8\pi (\eta \eta')^{1/2}\,\Delta}~\big[q\eta\eta']^\Delta \nonumber \\   K_{cc}[q,\eta,\eta'] &  = &
         -\frac{i}{8\pi^2}  {e^{-iq(\eta-\eta')}}\,\mathcal{P}\Big[\frac{1}{\eta-\eta'} \Big] + \, \frac{1}{8\pi}\,\delta(\eta-\eta')\,, \label{kernelmincoup}   \eea  The first line corresponds to the infrared enhancement of a minimally coupled, \emph{nearly massless} scalar field, arising from the infrared contribution of \emph{super Hubble} modes of the environmental scalar field  in the integral in (\ref{kernels}). The second line is recognized as the kernel for conformally coupled massless fields. Inspection of the different contributions available in ref.\cite{boyan} reveals the former arises from the regions $k\simeq 0~;~ |\vk+\vq| \simeq 0$ in the integral (\ref{kernels}) whereas the latter contribution    arises precisely from the terms $e^{-ip\eta}/\sqrt{2p}$ in the mode functions $u(p;\eta)$.

         With this kernel we now must obtain the non-local contribution to the quantum master equation   from $K_{IR}[q,\eta,\eta']$, in particular the coefficients of the respective terms in (\ref{rhoNLQP}), namely 
         \be \int^{\eta}_{\eta_0}   \,   \frac{g_{\pm}(q,\eta')}{\eta'} \, K_{IR}[q,\eta,\eta'] \, d\eta'\,.\label{IRcofs} \ee
         Following the arguments leading to (\ref{gpmdef}), the corresponding integrals are dominated by the region $-q\eta' \ll 1$, therefore focusing on this region and in the long time limit it follows that the first line in (\ref{kernelmincoup}) yields to leading order in $\Delta$
          \be K_{IR}[q,\eta,\eta']  \simeq \frac{(q^2\eta\eta')^{2\Delta}}{4\pi^2q^3(\eta\eta')^2\Delta}\,. \label{KIR} \ee With the super-Hubble behavior of $g_\pm$ given by (\ref{gis}) we find to leading order in $\Delta$
          \bea &&  \int^{\eta}_{\eta_0}   \,   \frac{g_{+}(q,\eta')}{\eta'} \, K_{IR}[q,\eta,\eta'] \, d\eta' \simeq \frac{1}{q^{3/2}\eta^2}\,\frac{1}{12\pi^2(-q\eta)^3\,\Delta}\nonumber \\ &&  \int^{\eta}_{\eta_0}   \,   \frac{g_{-}(q,\eta')}{\eta'} \, K_{IR}[q,\eta,\eta'] \, d\eta' \simeq \frac{1}{ q^{3/2}\eta^2}\,\frac{\ln[\eta/\eta^*]}{12\pi^2\,\Delta} \,. \label{gtils}\eea
          
          It becomes clear that the  coefficients of the $Q,P$ terms in the non-local contribution to the quantum master equation   now feature a much stronger secular contribution in the long time limit $\eta \rightarrow 0$ and are infrared enhanced by the pole in $\Delta$.


        In this case we expect that the corrections to the power spectrum of the ``system'' field will feature a stronger decay as discussed in the second reference in \cite{boyan}. This case merits further analysis and will be relegated to future study, as our goal in this article is to explore the impact of   environmental degrees of freedom with sub-Hubble correlations all throughout inflation.
         
         In refs.\cite{kahya} it was found that self-interactions of curvature perturbations and the interaction between a massless minimally coupled scalar field and curvature perturbations lead to time dependent logarithmic corrections (and powers of logarithms) to the power spectrum of the $\zeta$ (related to the curvature) variable, confirming previous results in ref.\cite{sloth1}. Although we did not study the curvature perturbation and we considered the inflaton-like scalar coupled to a massless conformal scalar field, our conclusions support in a non-perturbative manner the perturbative results of ref.\cite{kahya} in that the power spectrum is indeed time dependent (as well as scale dependent)  after ``horizon crossing''. Furthermore, the quantum master equation provides a non-perturbative resummation of  secular self-energy corrections (loops) and describes the  asymptotic long time behavior of correlations well after ``horizon crossing''.

\end{itemize}

\section{Conclusions and further questions.}
We studied the dynamics of an effective field theory of two interacting scalar fields during inflation: a minimally coupled inflaton-like field $\phi$ which is taken as the system  and another scalar  field $\varphi$   as  an \emph{environmental} quantum field that is integrated out. We obtained the reduced density matrix for  the   field  $\phi$ by tracing out the degrees of freedom of the scalar field $\varphi$, the reduced density matrix obeys a \emph{quantum master equation} which we obtained up to one loop in the correlation functions of the $\varphi$ field. The quantum master equation describes the decay of the vacuum state, the production of particles and correlated pairs and quantum  entanglement between the   fluctuations of the $\phi$ and $\varphi$. When the fluctuations of the $\phi$ field become super-horizon, the quantum master equation describes quantum entanglement between the super-Hubble degrees of freedom of the system  and sub-Hubble degrees of freedom of the environment. Renormalization aspects emerge naturally in this formulation. The quantum master equation provides an effective non-perturbative description of the dynamics whose solution is a resummation of self-energy corrections (loops). Our main goal in this article is to study the effect of environmental degrees of freedom that remain sub-Hubble all throughout inflation upon the power spectrum of super-Hubble fluctuations of the inflaton. For this purpose we considered the environmental field $\varphi$ to be a conformally coupled massless field, as a \emph{proxy} for degrees of freedom whose mode functions correspond to fluctuations that remain sub-Hubble all throughout inflation and are not amplified.   In this case   the quantum master equation provides a non-perturbative resummation of secular Sudakov-type double logarithms in the asymptotic long time limit. From the quantum master equation we obtain the time evolution of the power spectrum for super-Hubble fluctuations. Even when the non-interacting theory features a scale invariant power spectrum, the non-perturbative resummation of environmental correlations (loops) leads to a breakdown of scale invariance, and a \emph{decay} of the power spectrum at long time. Super-Hubble inflaton fluctuations do \emph{not} freeze out but decay upon ``horizon crossing''.
However for weak coupling between the inflaton and the environmental degrees of freedom, if inflation lasts for only $\simeq 60$ e-folds, the corrections to the power spectrum would be just marginally relevant for cosmological observations. This important aspect notwithstanding, the effective field theory description based on the reduced density matrix and its quantum master equation furnishes a powerful non-perturbative framework to study the impact of sub-Hubble degrees of freedom upon the quantum correlations of inflationary perturbations of cosmological relevance today.

\vspace{2mm}

\textbf{Further questions:} The results obtained above within a simple model of interactions between inflaton fluctuations and other degrees of freedom that remain sub-Hubble suggest several possible avenues of study. For example in the theory of non-Gaussianity the bi-spectrum is a result of a cubic self-interaction of curvature fluctuations\cite{maldacena,komatsu}. In the local limit for a squeezed configuration of the momenta of the three fields, two of the momenta are large and one becomes small, when the small momentum (the shortest side in the triangle) becomes super-Hubble and the other two are sub-Hubble, the situation is akin to the case studied in this article. In this case the bi-spectrum describes quantum entanglement between sub and super-Hubble degrees of freedom. Tracing over the sub-Hubble fluctuations leads to a quadratic effective action for the super-Hubble fluctuations much in the same way as the effective action for inflaton fluctuations considered here. Therefore, it would be interesting to obtain the corresponding quantum master equation by tracing over the sub-Hubble curvature fluctuations and study the asymptotic long time evolution of the power spectrum to understand if it indeed freezes out or if there is a correction such as the decay as found here in the simpler scenario. We have commented on the case when the environmental fields are nearly massless and minimally coupled to gravity highlighting the enhanced infrared and secular contributions as discussed in ref.\cite{kahya}. Pursuing a deeper understanding of this case with the quantum master equation is certainly of interest within this context because curvature fluctuations are described by massless and minimally coupled scalar fields, albeit with derivative interactions. Another relevant case to  consider is that with the inflaton Yukawa coupled to fermionic degrees of freedom with masses $\ll H$. As discussed above the mode functions for these degrees of freedom are very similar to those of a conformally coupled scalar field (with the ensuing spinorial structure). In the standard model there is a large number of fermionic degrees of freedom, which suggests a large N (number of fermionic species) expansion which furnishes yet another non-perturbative resummation scheme. Tracing over these fermionic degrees of freedom yield loop self-energies for the inflaton fluctuations, and the reduced density matrix and its concomitant quantum master equation can be obtained following the methods and steps highlighted here. For large N one can resort to the case in which the Yukawa coupling $\propto 1/\sqrt{N}$ so that the one loop self energy becomes exact in the large N limit. The results of some of these studies will be reported elsewhere.

\acknowledgements  The author  thanks the N.S.F. for partial
support through grant PHY-1202227.



\end{document}